\documentclass[a4paper, 11pt]{article}

\usepackage{authblk}
\usepackage{lineno}
\linenumbers

\usepackage{amsmath, amssymb}
\usepackage{graphicx}
\usepackage[colorlinks=true, allcolors=blue]{hyperref}
\usepackage{tabularx}
\usepackage{adjustbox}
\usepackage{caption}
\usepackage{subcaption}
\usepackage{svg}

\usepackage{soul} 
\usepackage{cmap} 
\usepackage{hyperref} \hypersetup{pdfstartview={XYZ null null 1.00}} 
\usepackage{enumerate} 

\title{ForeSpeed: A real-world video dataset of CCTV cameras with different settings for vehicle speed estimation}
\author[1]{Massimo Iuliani}
\author[2]{Blake Sawyer}
\author[1]{Marco Fontani}
\author[1]{David Spreadborough} 
\author[1]{Martino Jerian}
\affil[1]{Amped Software, Padriciano 99, AREA Science Park, 34149 Trieste, Italy.}
\affil[2]{Amped Software USA, 18 Bridge Street, Unit 2A, Brooklyn, NY 11201, USA}
\affil[3]{Azienda "Smart Mobility S.r.l.", Torino}
\date{\today}

\begin{document}

\maketitle

\begin{abstract}

The need to estimate the speed of road vehicles has become increasingly important in the field of video forensics, particularly with the widespread deployment of CCTV cameras worldwide.
Despite the development of various approaches, the accuracy of forensic speed estimation from real-world footage remains highly dependent on several factors—including camera specifications, acquisition methods, spatial and temporal resolution, compression methods, and scene perspective, which can significantly influence performance.

In this paper, we introduce ForeSpeed, a comprehensive dataset designed to support the evaluation of speed estimation techniques in real-world scenarios using CCTV footage.
The dataset includes recordings of a vehicle traveling at known speeds, captured by three digital and three analog cameras from two distinct perspectives.
Real-world road metrics are provided to enable the restoration of the scene geometry.
Videos were stored with multiple compression factors and settings, to simulate real-world scenarios in which export procedures are not always performed according to forensic standards.
Overall, ForeSpeed, includes a collection of $322$ videos.

As a case study, we employed the ForeSpeed dataset to benchmark a speed estimation algorithm available in a commercial product (Amped FIVE).
Results demonstrate that while the method reliably estimates average speed across various conditions, its uncertainty range significantly increases when the scene involves strong perspective distortion.
The ForeSpeed dataset is publicly available to the forensic community, with the aim of facilitating the evaluation of current methodologies and inspiring the development of new, robust solutions tailored to collision investigation and forensic incident analysis.
\end{abstract}



\linenumbers

\section{Introduction}

The speed of a vehicle prior to a collision is a critical factor in reconstructing the dynamics of an accident and assigning responsibility among the involved parties.
In many incidents, the only available visual evidence comes from closed-circuit television (CCTV) cameras. However, the quality of these recordings often falls short of ideal conditions, posing significant challenges for accurate forensic analysis.

In the last decades, several methods have been designed for image-based speed estimation targeting various application fields \cite{fernandez2021vision}.
Several approaches are designed to work in a controlled environment (e.g., traffic analysis), where automated speed estimation is attempted \cite{afifah2019vehicle}, \cite{huang2018traffic}, \cite{dougan2010real},
\cite{celik2009solar}, \cite{zhiwei2007models}, \cite{schoepflin2003dynamic}.
Many techniques leverage geometric cues like reverse projection \cite{epstein2019determination}, \cite{edelman2010tracking}, \cite{cathey2005novel}, or other perspective-based properties that can be applied in certain constraints, e.g., when the the vehicle has a longitudinal trajectory relative to the camera \cite{costa2020car} or the plate structure can be used to assess the vehicle speed \cite{luvizon2014vehicle}, \cite{vakili2020single}.
Motion analysis across consecutive frames is another common strategy, exploiting visual-related features to gather information on a vehicle speed \cite{lin2015biologically}, \cite{lan2014vehicle}, \cite{qimin2014methodology}.
It was also highlighted how the speed detection accuracy can be improved with the availability of multiple cameras or stereo images  \cite{llorca2016two}, \cite{el2018vehicle}.
Other exploit perspective properties, like cross-ratio \cite{han2016car}, and are conceived in a forensic scenario with CCTV cameras \cite{wong2014application}, or implementing even solutions robust to challenging scenarios, including nighttime conditions and uncertain visual anchors \cite{yang2024study}, \cite{choi2023novel}.

Although numerous methods exist, applying them in real-world scenarios can be challenging, since surveillance cameras are typically not designed for precise speed analysis.
Their settings, such as low spatial and temporal resolutions, intrinsically limit the accuracy of assessments.
Additionally, camera lenses can introduce strong image distortion that must be corrected prior to the estimate.
Other complications include severe perspective angles or distant viewpoints, as well as compression artifacts that can introduce uncertainty in determining the exact vehicle position.

Amidst this complex interplay of variables, the forensic practitioner must estimate the vehicle’s speed with the highest possible accuracy, bearing full responsibility for the chosen method or line of reasoning applied in the case.
From this standpoint, the proper acquisition of video evidence represents the critical first step in preserving both recorded events and associated metadata \cite{lallie2023dashcam}, \cite{wooller2005digital}.
Subsequently, the collision investigator must apply an appropriate speed estimation technique, while also evaluating the reliability of the result within the specific case context, which is often far away from ideal conditions \cite{hoogeboom2010measurement}.
Indeed, officers frequently encounter material that has not been acquired in compliance with best practices \cite{swdge2024}, especially when it comes to the proper export of video files. 

A number of datasets have been introduced in the literature to evaluate speed estimation methods.
In some cases, these datasets focus on specific imaging devices, such as dashcams \cite{lallie2020dashcam}.
However, there remains limited experimental validation for scenarios that reflect the uncontrolled conditions typical of forensic investigations using CCTV footage.
One of the most widely used dataset for traffic and vehicle speed
detection is the AI City Challenge \cite{hua2018vehicle}, \cite{huang2018traffic}.
It provides a few video tracks recorded in normal traffic situations, such as highways or intersections.
Although the scene is captured with a very high meter-to-pixel ratio, the videos are captured with a high frame rate, and there is no variety of camera models. 
Similarly, BrnoCompSpeed \cite{sochor2017brnocompspeed} is a dataset that focuses on camera calibration for traffic analysis and visual speed measurements from a single monocular camera.
The dataset contains 18 videos at a resolution of 1920×1080 pixels at 50
fps, including more than 20,000 tagged vehicles.
However, the video is captured from three cameras (Panasonic HC-X920, Panasonic HDC-SD90, and Sony Handycam HDR-PJ410) with specifics that are very far from common CCTV cameras.
Another dataset provides hours of video recorded under varying weather conditions on urban roads.
Although having different weather conditions can be useful to assess its impact on the task, only a single camera is available \cite{luvizon2016video}.
A more forensic-oriented effort can be found in \cite{shim2021vehicle}, which explores speed estimation using dashcam videos recorded in diverse road environments.
In this work, the author also considered some uncomfortable scenarios, including working with retaken and re-encoded videos which may cause alterations of playback speed and time resolution.
However, no CCTV cameras are involved in the dataset.

In this paper, we introduce ForeSpeed,  a real-world video dataset of CCTV
cameras with different settings for the analysis and testing of vehicle speed estimation methods.
The dataset consists of video recordings of a vehicle traveling at known speeds, captured by three digital and three analog cameras combined to three different digital video recorders (DVRs) with varying technical specifications, including differences in resolution and encoding schemes.
The vehicle was filmed from two distinct viewpoints: one in which it moves parallel to the camera, and another where it approaches or recedes from the camera with stronger perspective while maintaining constant speed.

The dataset is specifically designed to analyze if and how individual parameters may influence the performance of speed estimation methods.
As a case study, we applied ForeSpeed to benchmark one such method.
Our analysis highlights that the method is effective in estimating the vehicle speed independently of most parameters.
However, its uncertainty level strongly increases in scenes with strong perspective.
ForeSpeed is publicly released to support the forensic community as a reference benchmark for validating other speed estimation techniques on CCTV footage.


\section{ForeSpeed: a dataset for forensic speed estimation}
\label{sec:data}

The proposed dataset serves as a benchmark for evaluating the impact of various image acquisition factors on the accuracy of speed estimation.
These factors include scene perspectives, camera models, their technical specifications, recording and exporting settings.

Scene perspective plays a critical role in both point selection and meter-to-pixel ratio, that is, the portion of the road segment represented by each pixel.
When the road is viewed from a low-angle perspective (where its direction is nearly parallel to the image plane), the meter-to-pixel ratio tends to remain relatively consistent along the vehicle’s path.
In contrast, a strong perspective, where the road direction forms a wide angle with the image plane, introduces significant variation in the vehicle’s distance from the camera as it moves.
This change has a notable impact on pixel interpretation: as the vehicle moves away from the camera, the meter-to-pixel ratio increases rapidly, and identifying the contact points between the wheels and the road becomes increasingly uncertain. 

Different lens models and encoding schemes can produce markedly different digital representations of a video sequence.
Lens distortion may cause straight lines to appear curved, compromising geometric measurements, while heavy compression can blur edges, introducing a laundering effect that complicates accurate point selection.
These factors can significantly influence the speed estimation process and must be appropriately addressed: lens distortion should be corrected, and uncertainty in point selection should be evaluated on a case-by-case basis by the forensic analyst.
In addition, the level of details, when recording the same scene, strongly varies depending on the camera's spatial and temporal resolution.
For example, low-resolution imagery and variable frame rate acquisition can substantially degrade the precision of the speed estimate.
%
\begin{figure}
\centering
\begin{subfigure}{.5\textwidth}
  \centering
\includegraphics[width=.9\linewidth]{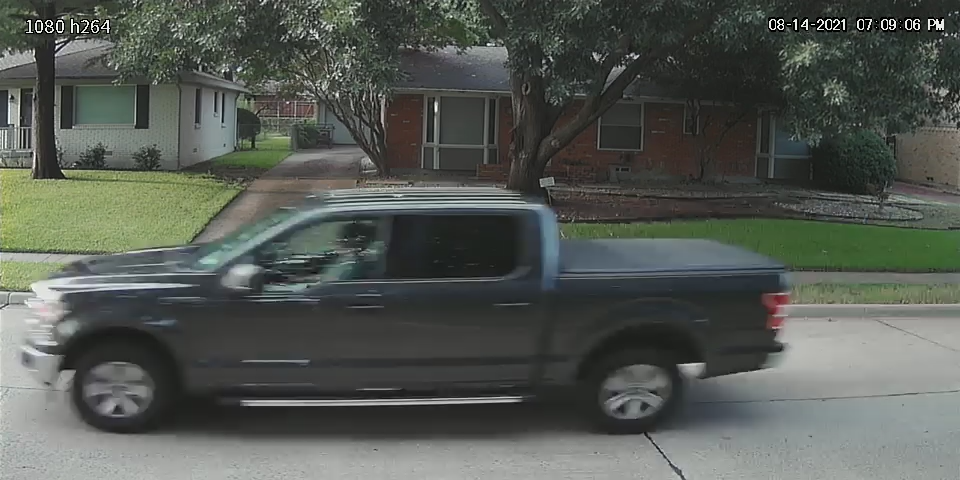}
  \caption{Low perspective view.}
  \label{fig:sub1}
\end{subfigure}%
\begin{subfigure}{.5\textwidth}
  \centering
\includegraphics[width=.66\linewidth]{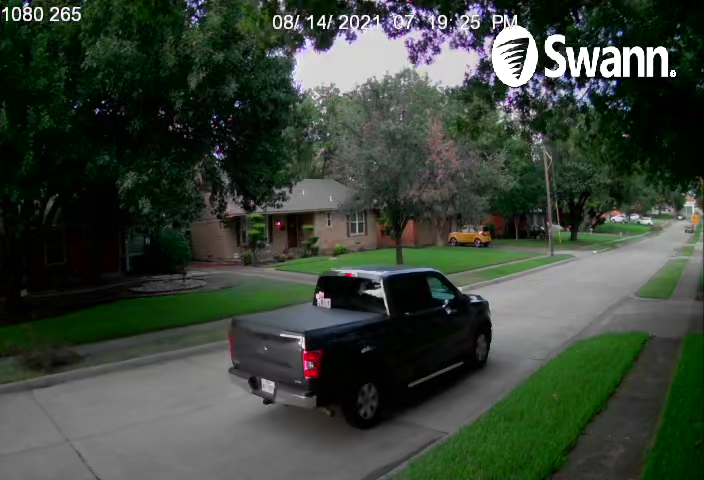}
  \caption{Strong perspective view.}
  \label{fig:sub2}
\end{subfigure}
\caption{Example of different scene perspectives.}
\label{fig:persp}
\end{figure}

\begin{figure}
\centering
\includegraphics[width=\linewidth]{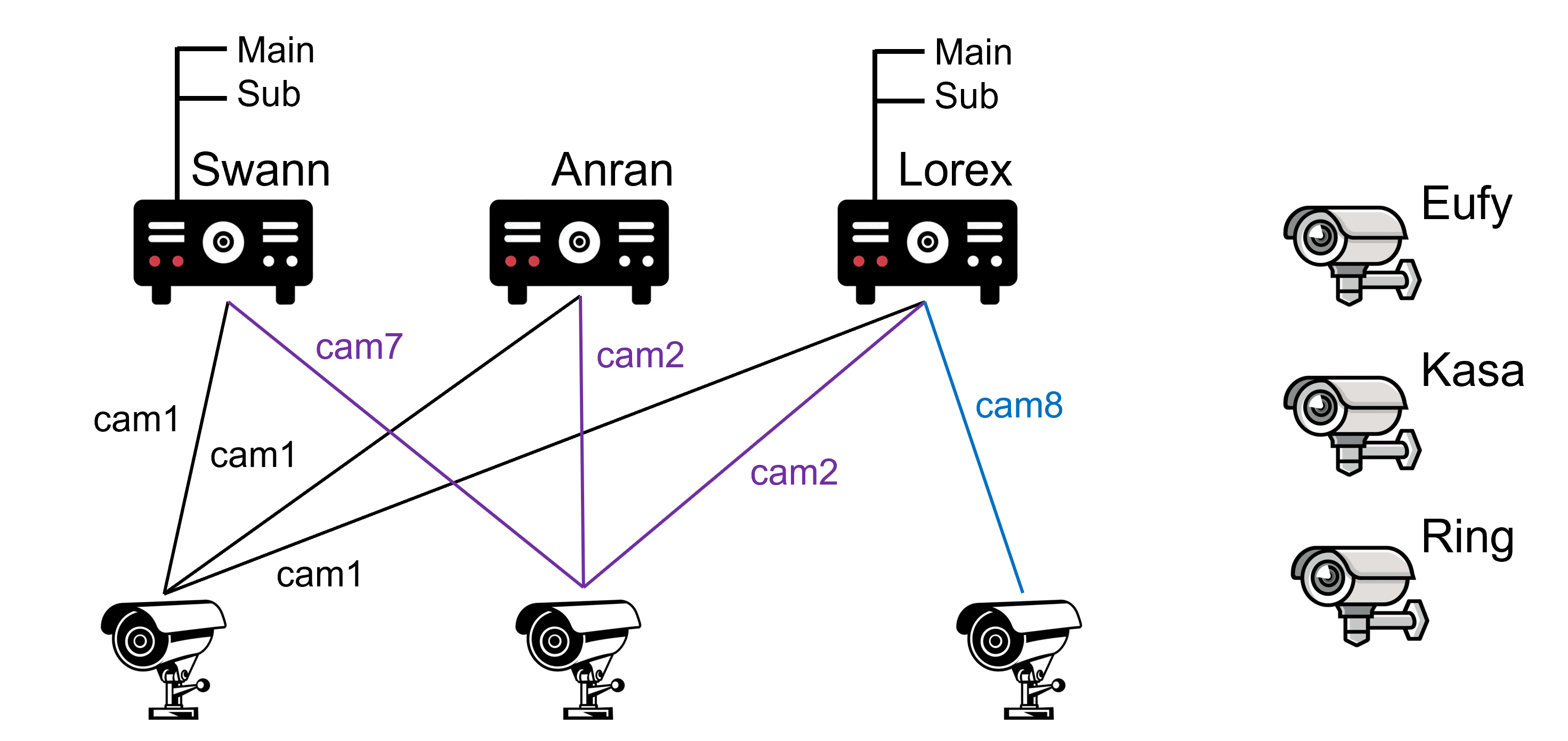}
\caption{Dataset structure: three digital cameras and three analog cameras connected to different DVRs.}
\label{fig:7cameras}
\end{figure}

To account for such variability, we employed both digital and analog cameras, the latter connected to digital video recorders, as in the following:
\begin{itemize}
    \item three digital cameras, namely Eufy (T8441), Kasa (KC105), Ring (2nd generation);
    \item two 1080p analog cameras, each connected to three DVRs, namely Anran, Lorex (D862A8-Z), Swann (DVR8-4680WRN);
    \item a 4k analog camera connected to the Lorex DVR (Anran and Swann did not support such a resolution).
\end{itemize}
Both Lorex and Swann allowed to acquire two different streams, namely Main and Sub, both reflected in the dataset.
Moreover, when possible, multiple formats were exported (.avi and .mp4).
A diagram of the system used is provided in Figure \ref{fig:7cameras}.

The above setup was used to capture a vehicle traveling on a road at a known speed.
All cameras were mounted on a wooden platform, enabling them to capture the scene from approximately the same position.
The platform was repositioned to record from two distinct perspectives: (i) a low perspective view, where the vehicle runs nearly parallel to the image plane; and (ii) a strong perspective view, where the road direction forms a wider angle relative to the image plane.
Figure \ref{fig:persp} presents an example of the same scene captured from both perspectives.

\begin{figure}
\centering
\begin{subfigure}{.5\textwidth}
  \centering
\includegraphics[width=.8\linewidth]{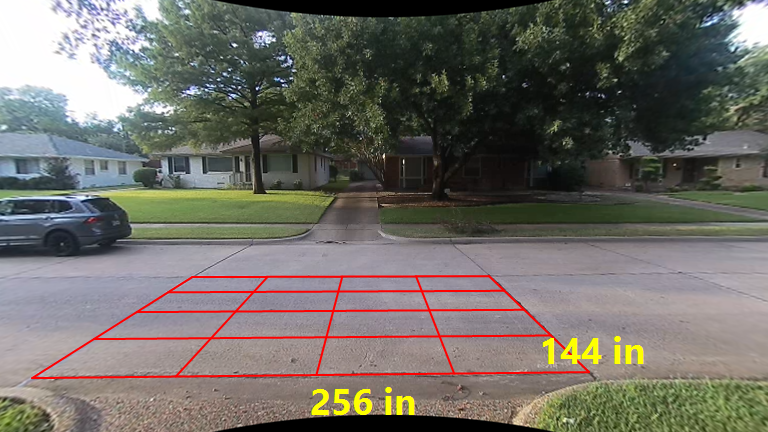}
  \caption{Low perspective case}
  \label{fig:meas1}
\end{subfigure}%
\begin{subfigure}{.5\textwidth}
  \centering
\includegraphics[width=.8\linewidth]{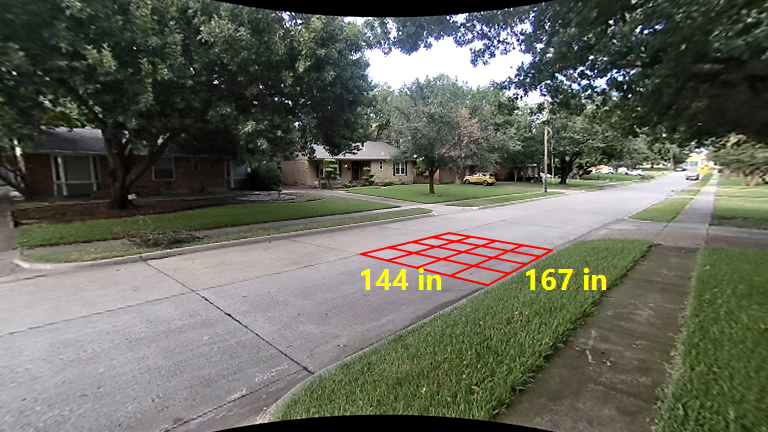}
  \caption{Strong perspective case}
  \label{fig:meas2}
\end{subfigure}
\caption{Reference measures}
\label{fig:ref_measures}
\end{figure}

\begin{figure}
\centering
\includegraphics[width=0.50\linewidth]{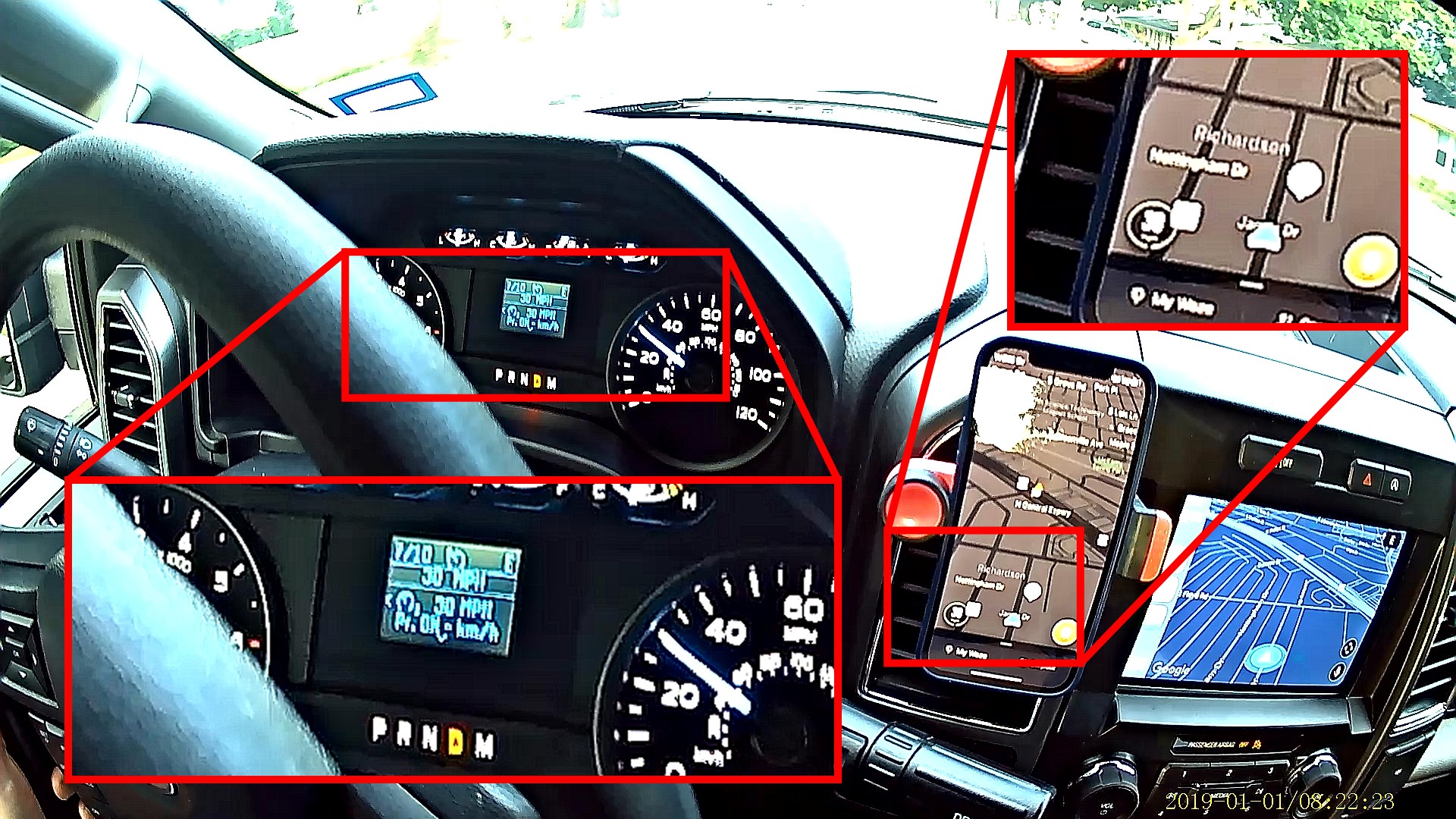}
\caption{A camera captured the speed visualized on the speedometer and an external GPS-based application.}
\label{fig:GTspeed}
\end{figure}

More specifically, we collected $6$ and $8$ passes for the low and strong perspective cases, respectively.
The vehicle, a 2020 Ford F-150 4x2 Supercrew\footnote{Overall, the vehicle is $231.9$ inches, with a wheelbase of 145 inches. Technical specs of the vehicle can be found on \url{https://media.ford.com/content/dam/fordmedia/North\%20America/US/product/2020/f150/2020-F150-TechSpecs.pdf}}, passes at $30$ mph.
The exact speed was checked during every pass through the combined visualization of the car speedometer and an independent GPS-based application.
Both speeds were captured by an independent camera during each pass, as shown in Figure \ref{fig:GTspeed}.
The car speedometer was also independently validated through a laser speed sensor to verify its reliability.
The speed of each pass is reported in Table \ref{tab:GT}.
Each pass was simultaneously acquired with all available cameras and stored in multiple formats, when possible.
Furthermore, we collected some measures from the road that can be used as a reference for estimating the vehicle speed.
We report such values in Figure \ref{fig:ref_measures}.

Since ForeSpeed is designed to resemble real-world scenarios, we exported each video with various formats and settings, based on the methods available on each camera.
The full list of available camera settings and specifications is reported in Table \ref{tab:camera_settings}.
For each video, we report the acquisition camera, the spatial resolution, the observed codec, the storage format, the frame rate, and the type of frame rate (constant or variable).
Overall, ForeSpeed includes $89$ and $180$ videos under low and strong perspectives, respectively, for a total of $269$ videos.

\begin{table}[]
\caption{List of camera acquisition models, settings, and formats (vfr stands for variable frame rate).}
\label{tab:camera_settings}
\begin{center}
\begin{adjustbox}{width=0.8\textwidth}
\begin{tabular}{lllllll}
Model & Camera    & Format & Codec & Resolution &  Fps   \\
\hline
Anran & Cam1      & avi    & hevc  & 1920x1080  & 14    \\
Anran & Cam2      & avi    & hevc  & 1920x1080  & 14    \\
Eufy  & Cam1      & mp4    & h264  & 2304x1296  & 14.99 (vfr) \\
Kasa  & Cam1      & mp4    & h264  & 1920x1080  & 14.99 (vfr) \\
Lorex & Cam1 main & avi    & hevc  & 1920x1080  & 30 \\
Lorex & Cam1 main & mp4    & hevc  & 1920x1080  & 30   \\
Lorex & Cam1 sub  & mp4    & hevc  & 704x408    & 15 (vfr)    \\
Lorex & Cam2 main & avi    & h264  & 1920x1080  & 30   \\
Lorex & Cam2 main & mp4    & h264  & 1920x1080  & 30   \\
Lorex & Cam2 sub  & mp4    & h264  & 960x480    & 10    \\
Lorex & Cam8 main & avi    & hevc  & 3840x2160  & 7    \\
Lorex & Cam8 main & mp4    & hevc  & 3840x2160  & 7   \\
Lorex & Cam8 sub  & mp4    & h264  & 960x480    & 7   \\
Ring  & DevTools  & mp4    & h264  & 640x360    & 10   \\
Ring  & Download  & mp4    & h264  & 1920x1080  & 10    \\
Swann & Cam1 main & avi    & hevc  & 1920x1080  & 15    \\
Swann & Cam1 main & mp4    & hevc  & 1920x1080  & 15.03 (vfr) \\
Swann & Cam1 sub  & avi    & hevc  & 704x480    & 5     \\
Swann & Cam1 sub  & mp4    & hevc  & 704x480    & 5 (vfr)     \\
Swann & Cam7 main & avi    & h264  & 1920x1080  & 15    \\
Swann & Cam7 main & mp4    & h264  & 1920x1080  & 15.03 (vfr) \\
Swann & Cam7 sub  & avi    & h264  & 704x480    & 5     \\
Swann & Cam7 sub  & mp4    & h264  & 704x480    & 5.02 (vfr)  \\
                                        \hline
\end{tabular}
\end{adjustbox}
\end{center}
\end{table}

\section{2D Speed Estimation}
\label{sec:speed_est}
In this section, we summarize the speed estimation algorithm implemented in Amped FIVE that will be validated on ForeSpeed.

\begin{figure}
\centering
\begin{subfigure}{.5\textwidth}
  \centering
\includegraphics[width=.8\linewidth]{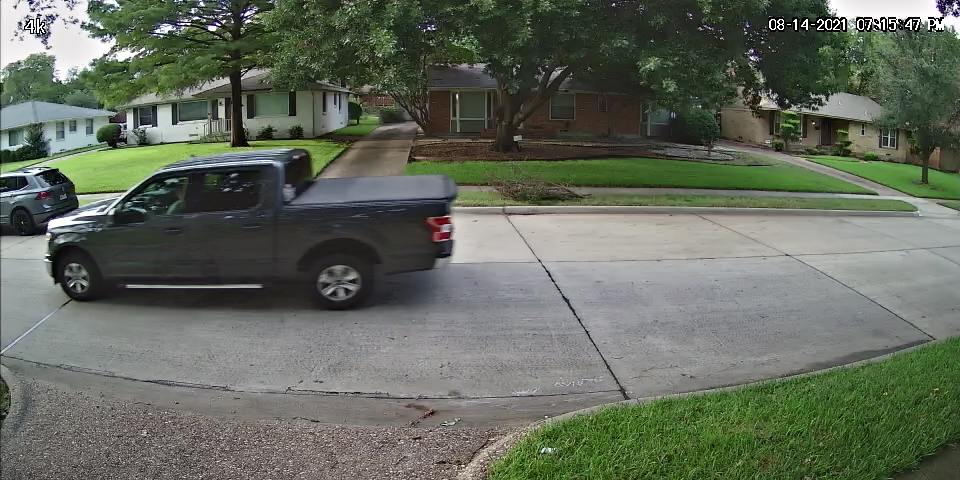}
  \caption{Image acquired by the camera}
  \label{fig:distort}
\end{subfigure}%
\begin{subfigure}{.5\textwidth}
  \centering
\includegraphics[width=.8\linewidth]{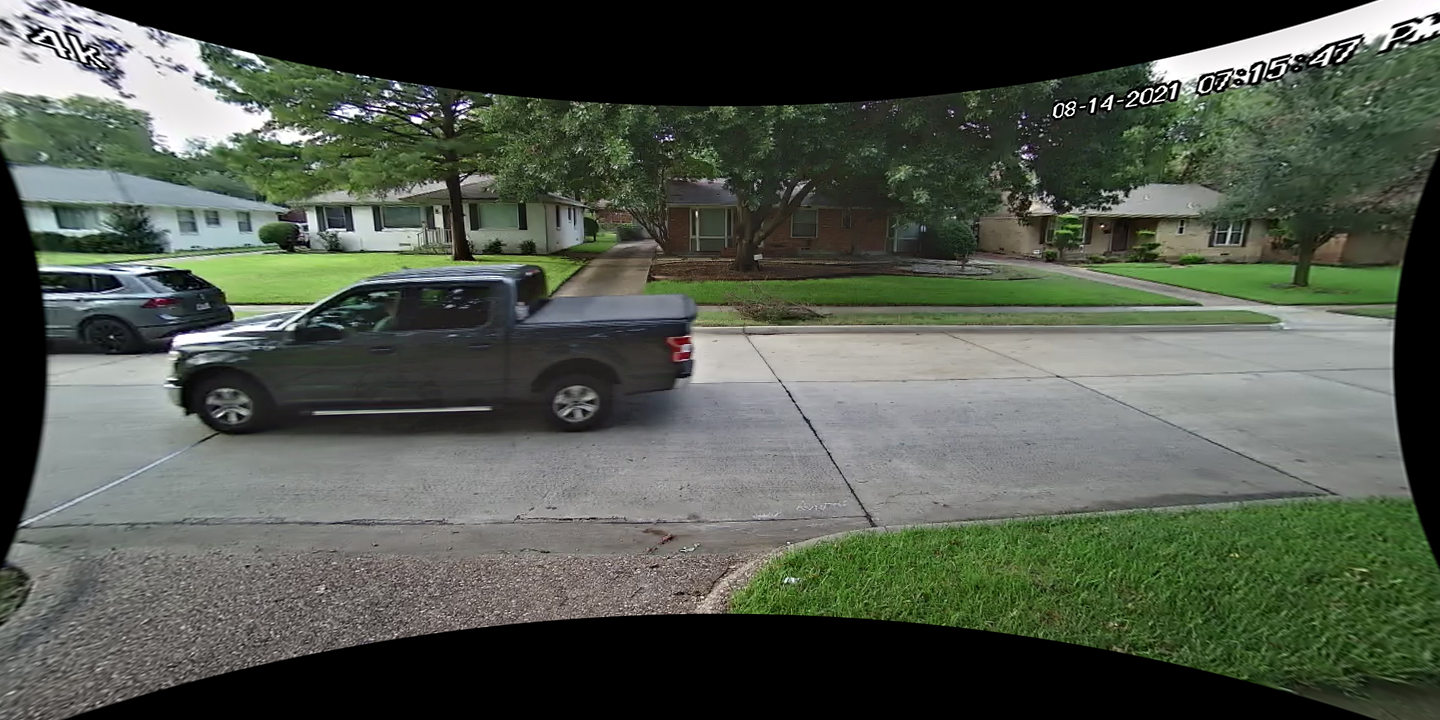}
  \caption{Image with corrected distortion}
  \label{fig:undistort}
\end{subfigure}
\caption{Example lens distortion removal}
\label{fig:dist_undistort}
\end{figure}

As an initial processing step, any optical distortion in the image must be corrected.
This correction relies on the principle that straight lines in the real-world should also appear straight in the 2D image.
The distortion can be estimated by manually selecting the curved lines in the image that are expected to be straight once the distortion has been corrected \cite{de1998deinterlacing}.
We report an example of an original frame, as acquired by the camera, and the corresponding frame after correcting the lens distortion in Figure \ref{fig:dist_undistort}.

Under the assumption of a flat road, the real-world distances can be computed exploiting the actual dimensions of a rectangular object on the road:
the four corners of the object can be used for image rectification \cite{jain1989fundamentals}, thus removing the perspective distortion.
The known size of the rectangular object can be directly exploited to remove the unknown scale factor and compute the real-world distances.
%
\begin{figure}
\centering
\begin{subfigure}{.45\textwidth}
  \centering
\includegraphics[width=\linewidth]{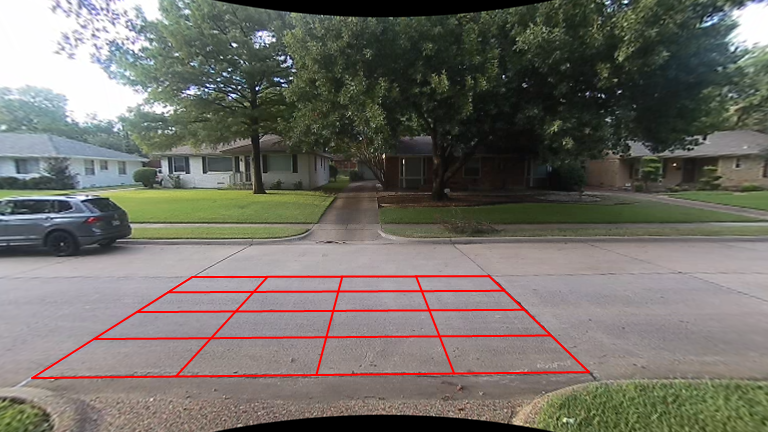}
  \caption{Grid selection.}
  \label{fig:m1}
\end{subfigure}%
\begin{subfigure}{.44\textwidth}
  \centering
\includegraphics[width=\linewidth]{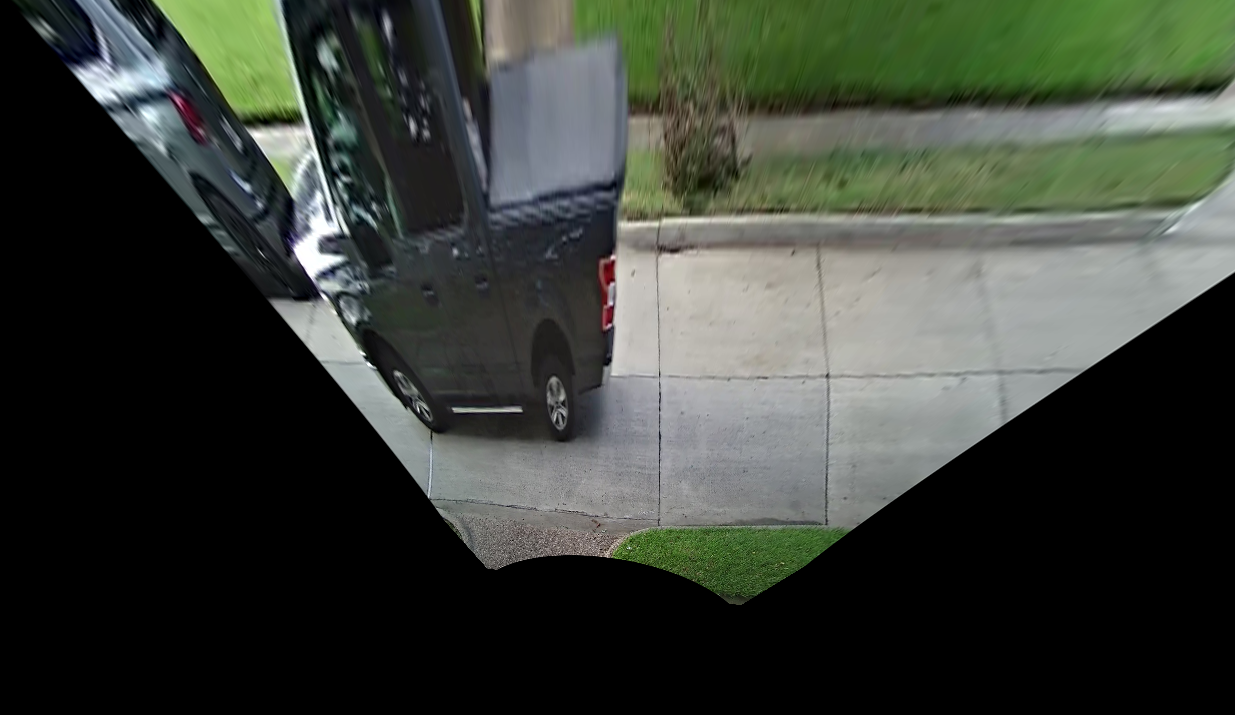}
  \caption{Road rectification.}
  \label{fig:m2}
\end{subfigure}
\caption{Example of reference selection and road rectification.}
\label{fig:rect}
\end{figure}
In Figure \ref{fig:rect}, we report an example of a grid selection and the aerial image obtained by applying a reverse projection on the selected grid.

The vehicle's path can be defined as a sequence of contact points (CPs) between the tires and the road.
%
\begin{figure}
\centering
\begin{subfigure}{.5\textwidth}
  \centering
\includegraphics[width=.8\linewidth]{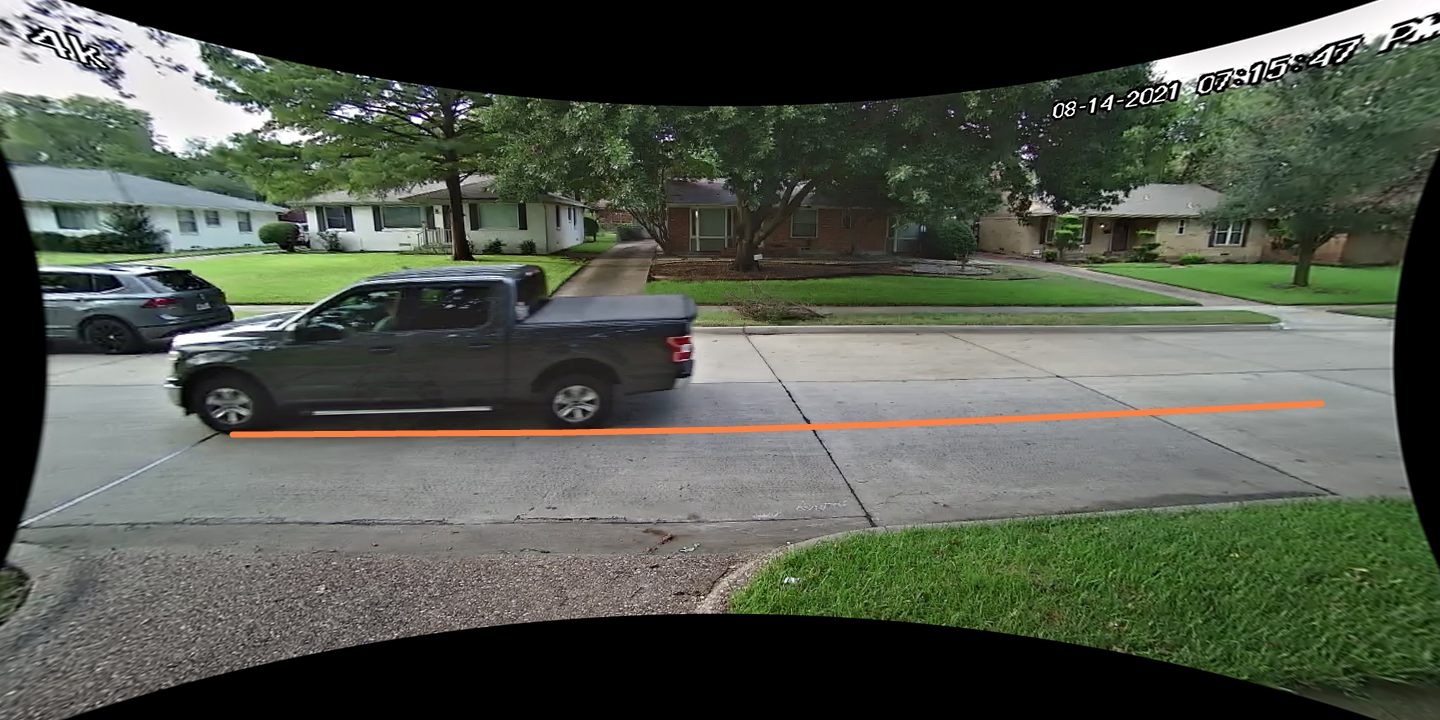}
  \caption{Example of selected path}
  \label{fig:path}
\end{subfigure}%
\begin{subfigure}{.5\textwidth}
  \centering
\includegraphics[width=.64\linewidth]{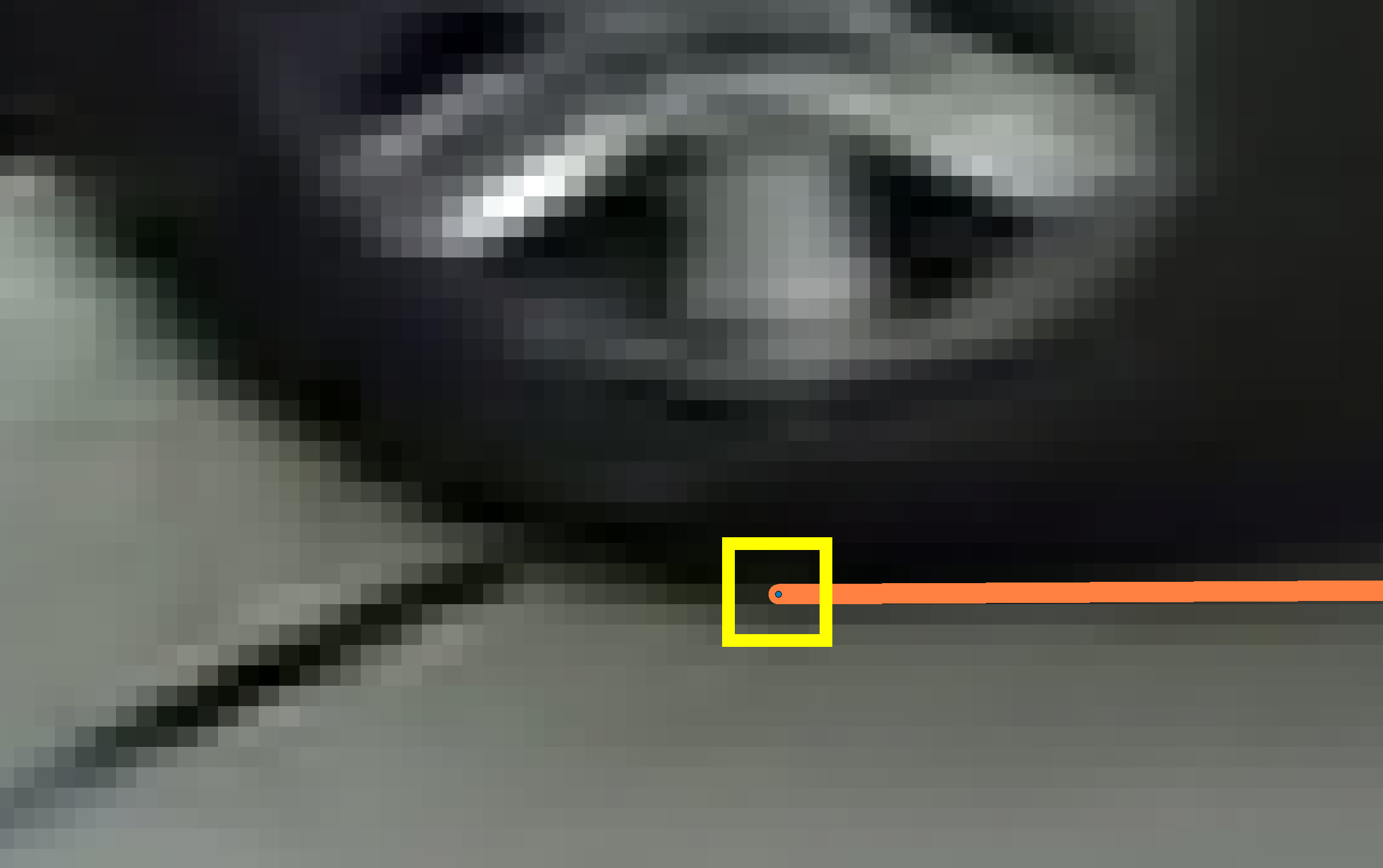}
  \caption{Example of contact point (CP) uncertainty}
  \label{fig:CP_uncertainty}
\end{subfigure}
\caption{Example of path and contact points selection}
\label{fig:path_and_uncert}
\end{figure}
When defining the $i$-th CP, the user can also set a pixel uncertainty of $m_i$ pixels, defining an error range of $(2m_i+1)\times (2m_i+1)$ pixels in which the actual CP is expected to be (see example in Figure \ref{fig:path_and_uncert}).

The geometric correction of the image allows to obtain the rectified CPs and uncertainty regions.
\begin{figure}
\centering
\includegraphics[width=0.80\linewidth]{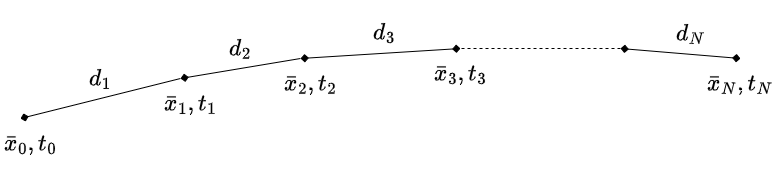}
\caption{List of contact points defining the vehicle path.}
\label{fig:path_modelling}
\end{figure}
Formally, we can define the rectified path as a sequence of rectified CPs ($\bar{x}_i,t_i$), where $\bar{x}_i$ are the rectified spatial coordinates and $t_i$ are the temporal coordinates for each frame $i$, with $i=0, \dots, N$. 
The path defines a trajectory of length $d$, consisting of $N$ segments of length $d_1,\dots, d_N$, where $d_i$ is the distance between $\bar{x}_{i-1}$ and $\bar{x}_{i}$ (see Figure \ref{fig:path_modelling}).

\begin{figure}[h]
	\centering
	\includegraphics[scale=0.3]{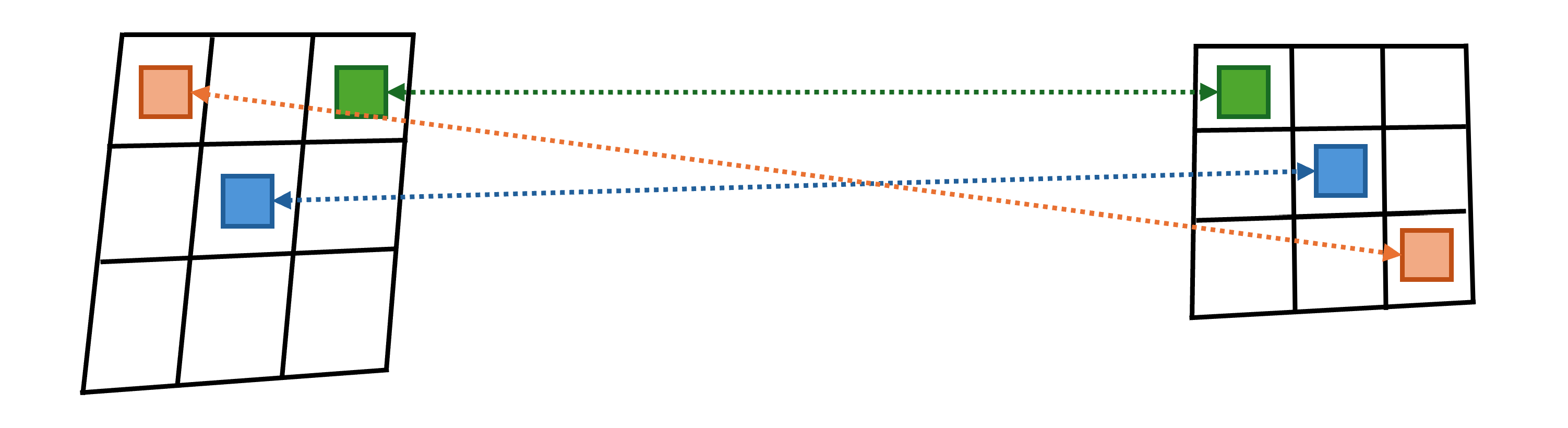}
	\caption{Example of rectified CPs in two consecutive positions with a pixel uncertainty $m=1$. The selected distance $d_i$(the blue segment) is comprised between the maximum distance $d^{(max)}_{i}$ (the orange segment) and the minimum distance $d^{(min)}_{i}$ (the green segment).}
	\label{fig:CP_undertainty}
\end{figure}
The distance $d_i$ (blue arrow) is defined as the distance between the two rectified CPs (the two blue squares in Figure \ref{fig:CP_undertainty}).
We denote with $d^{(max)}_i$ and $d^{(min)}_i$ the maximum and minimum distances achievable by connecting points of the first rectified uncertainty region to points of the second rectified uncertainty region. In the figure, we visually show the maximum distance $d^{(max)}_i$ (the orange segment) and the minimum distance $d^{(min)}_i$ (the green segment) for that case.

Formally, we can define the error $\Delta d_i$ as
\begin{equation}
\label{eq:delta_d}
    \Delta d_i := max\{d^{(max)}_{i} - d_i, d_i - d^{(min)}_{i}\}.
\end{equation}

In the end, a conservative definition of the overall distance error can be defined by summing the distance error on each segment:
\begin{equation}
\Delta d= \sum_{j=1}^{N} \Delta d_j, \hspace{1cm} \mbox{with} \hspace{1cm} d= \sum_{j=1}^{N} d_j .
\end{equation}
The time uncertainty can be computed differently since it is mainly affected by $t_0$ and $t_N$.
For simplicity, we assume that $\Delta t_0$ = $\Delta t_N$ = $\Delta t$.
The timing source from CCTV can often be obtained from the video’s presentation timestamps (PTS), or the CCTV data timestamp allocated to the video frame that represents date and time.
Both of these though require evaluation to ensure integrity, authenticity and reliability. Timing and then the duration of each frame can also be calculated from the number of frames and duration; however, this will be an average and will not take variability into account.
An external time reference, e.g.,a lightboard, can be exploited to assist in establishing variability and aid in the setting of an objective error rate to the individual frame times \cite{crouch2017use}.

The average speed of the vehicle can be computed as
\begin{equation}
v = \frac{d}{t_N-t_0}
\end{equation}
with relative distance ($\epsilon_d$) and time ($\epsilon_t$) errors
\begin{equation}
\epsilon_d = \frac{\Delta d}{d} \hspace{2cm} \epsilon_t = \frac{2 \cdot \Delta t}{t_f-t_i}.
\label{eq:rel_err}
\end{equation}
Then, the relative speed error can be obtained as $\epsilon_v = \epsilon_d + \epsilon_t$.
Finally, the absolute speed error can be trivially obtained as 
\begin{equation}
\Delta v = \epsilon_v \cdot v.  
\label{eq:dv}
\end{equation}
In conclusion, starting from the user-defined uncertainty $\Delta t$ and $\Delta d_j$, with $j=1, \dots, N$, we can determine the average speed $v$ and the speed value range $[v - \Delta v, v + \Delta v]$.
Note that the above method, contrary to the cross-ratio approach \cite{wong2014application, han2016car}, allows speed computation even when the vehicle is turning since the user can define a segmented path.
In the most common cases, however, under the assumption of a straight path, we can simplify the modeling of the path with two points only: the starting point ($\bar{x}_0,t_0$) and the final point ($\bar{x}_1,t_1$).
\paragraph{Case Study: NFI Collaborative Exercise}
The described speed estimation method was exploited in the ``Collaborative Exercise of Speed from Video 2023'', proposed by the Netherlands Forensic Institute and distributed among the ENFSI Digital Imaging Working Group and Road Accident Analysis Working Group.
Given a questioned video file, examiners were asked to determine the average speed of a car over a given trajectory.
The questioned video had good quality, with a spatial resolution of $3840\times2160$ and $15$ frames per second.
Car characteristics, real-world measurements, and other reference recordings were also provided.
In this specific case, the path is composed by two road segments on different planes connected by a bump.
For that reason, the speed was estimated with two measurements, before and after the bump, respectively (namely, Measure 1 and Measure 2).
In both cases grid measurements from the road were used to get real-world distances.
We achieved $82.5 \pm 8.5$ and $84.4 \pm 8.8$ Km/h respectively, over a real car speed of $81$ Km/h, thus overestimating the average speed of $1.01 \%$ and $1.04 \%$ respectively.

It is worth analyzing the uncertainty estimation in this specific case, as each car’s trajectory includes only five frames.
Short distances and time intervals inherently increase the relative speed error (see Equation \ref{eq:rel_err}), regardless of the method used.
To highlight the significance of this factor, Table \ref{tab:speed_unc} reports the absolute speed error obtained in the two measurements, based on the length of the considered path.
The first column shows the evaluated path length (for example, d1 + d2 indicates that speed is computed over the distance between frames 1 and 3). As expected, the more frames included in the computation, the lower the resulting uncertainty and the lowest uncertainty is observed when all available frames are used.
This example demonstrates that limited information can significantly affect uncertainty, even when the estimated speed closely matches the actual value.
%
\begin{table}[]
\begin{center}
\begin{tabular}{c|c|c}
Measure 1           & $v$           & $\Delta v$    \\ \hline
$d_1$               & $82.0$        & \textbf{26.0} \\ \hline
$d_1+d_2$           & $82.6$        & \textbf{14.1}  \\ \hline
$d_1+d_2+d_3$       & $83.0$        & \textbf{10.3}  \\ \hline
$d_1+d_2+d_3+d_4$   & \textbf{82.5} & \textbf{8.6}  
\end{tabular}
\quad
\begin{tabular}{c|c|c}
Measure 2           & $v$           & $\Delta v$ \\ \hline
$d_1$               & $81.8$        & \textbf{31.8}    \\ \hline
$d_1+d_2$           & $84.2$        & \textbf{16.7}  \\ \hline
$d_1+d_2+d_3$       & $84.5$        & \textbf{11.4}  \\ \hline
$d_1+d_2+d_3+d_4$   & \textbf{84.4} & \textbf{8.8}  
\end{tabular}
\caption{Speed estimation achieved in the considered case study. Two independent estimations were performed (Measure 1 and 2) based on different real-world measurements. We assess the speed on different path lengths.
For instance, $d_1+d_2$ is referred to the speed computed on the first two distance segments (between frames 1 and 3).}
\label{tab:speed_unc}
\end{center}
\end{table}

\section{Experimental Results}
\label{sec:exp}
The proposed dataset was employed to evaluate the reliability of the speed estimation method described in Section \ref{sec:speed_est} across different scene perspectives, camera models, configurations, and exporting settings.
Emulating a forensic scenario, the method was applied manually to each pass captured by each camera. 
Lens distortion correction and speed estimation was performed using the implementation provided by Amped FIVE.

For each video sequence, the user selected the longest visible segment of the vehicle’s path.
Videos for which contact points were not clearly visible throughout the path were excluded from the analysis.
Uncertainty boxes were defined to ensure they contained the corresponding contact points (see Figure \ref{fig:CP_uncertainty}).
The time uncertainty was set to the default value $\Delta t = 5$ ms.  
In the scene, the known dimensions of the rectangular ground grid in the scene were manually selected to rectify the road plane and estimate real-world distances.
For each video we achieved an average speed $v$ and an uncertainty range $\Delta v$ (as defined in Equation \ref{eq:dv}).
Since the vehicle is moving in a straight line, we considered the longest visible path with two CPs only.

We then examined the error of the estimated average speed with respect to the ground-truth value. 

The analysis led to the following key findings: the estimation error for $v$, averaged across all experiments, is $-0.96$ mph. Considering that the speedometer resolution is 1mph, this value indicates the absence of a strong bias in the estimation method, possibly suggesting a slight tendency to underestimate the speed. 
If we split the results according to the two scene perspectives, low and strong, the average speed errors are 
$-0.93$ mph and $-0.99$ mph, respectively.
This suggests that the scene perspective has minimal influence on the average performance of the system.

\begin{figure}
\centering
\begin{subfigure}{.5\textwidth}
  \centering
\includegraphics[width=\linewidth]{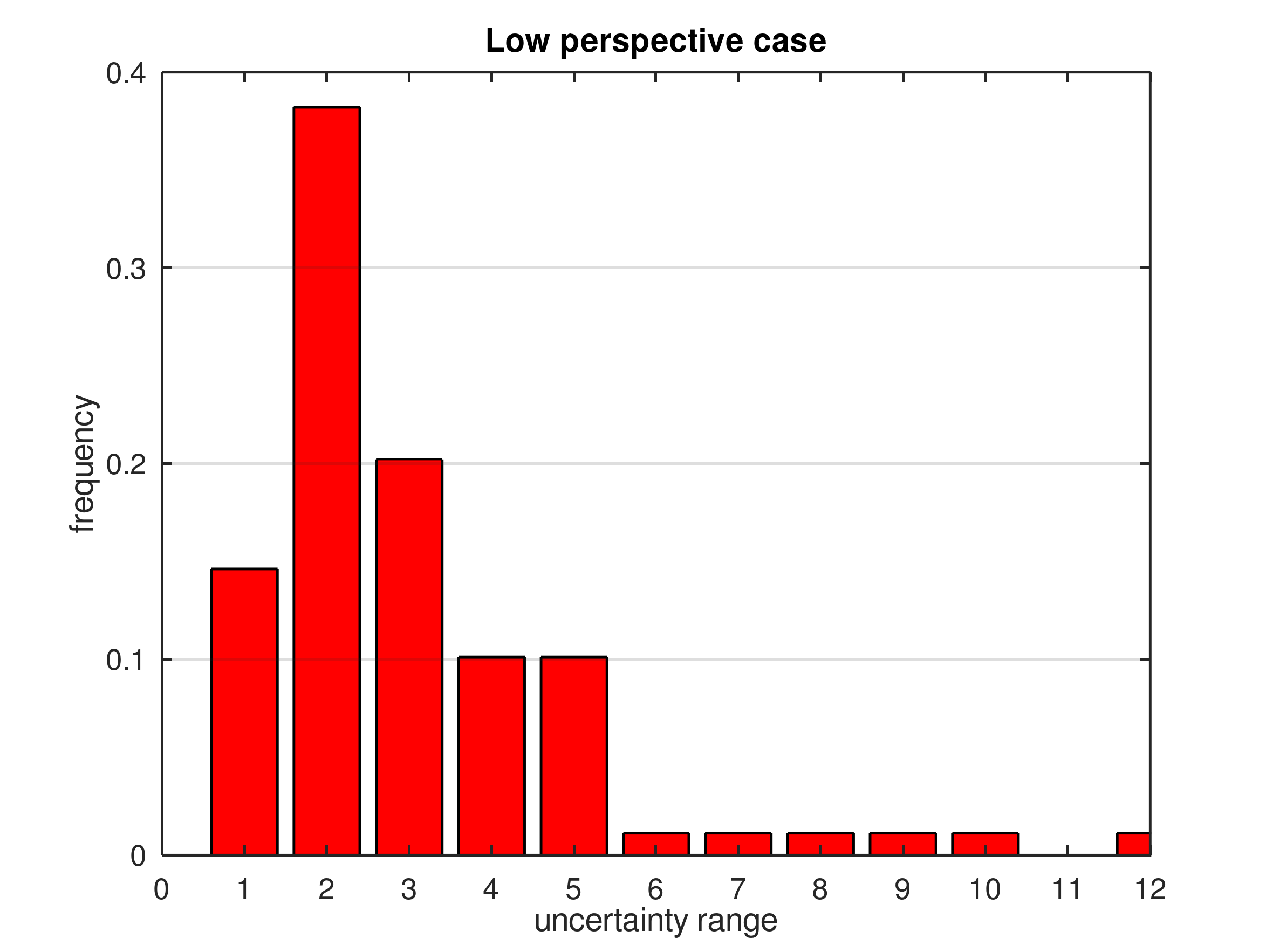}
  \caption{Low perspective case}
  \label{fig:h1delta}
\end{subfigure}%
\begin{subfigure}{.5\textwidth}
  \centering
\includegraphics[width=\linewidth]{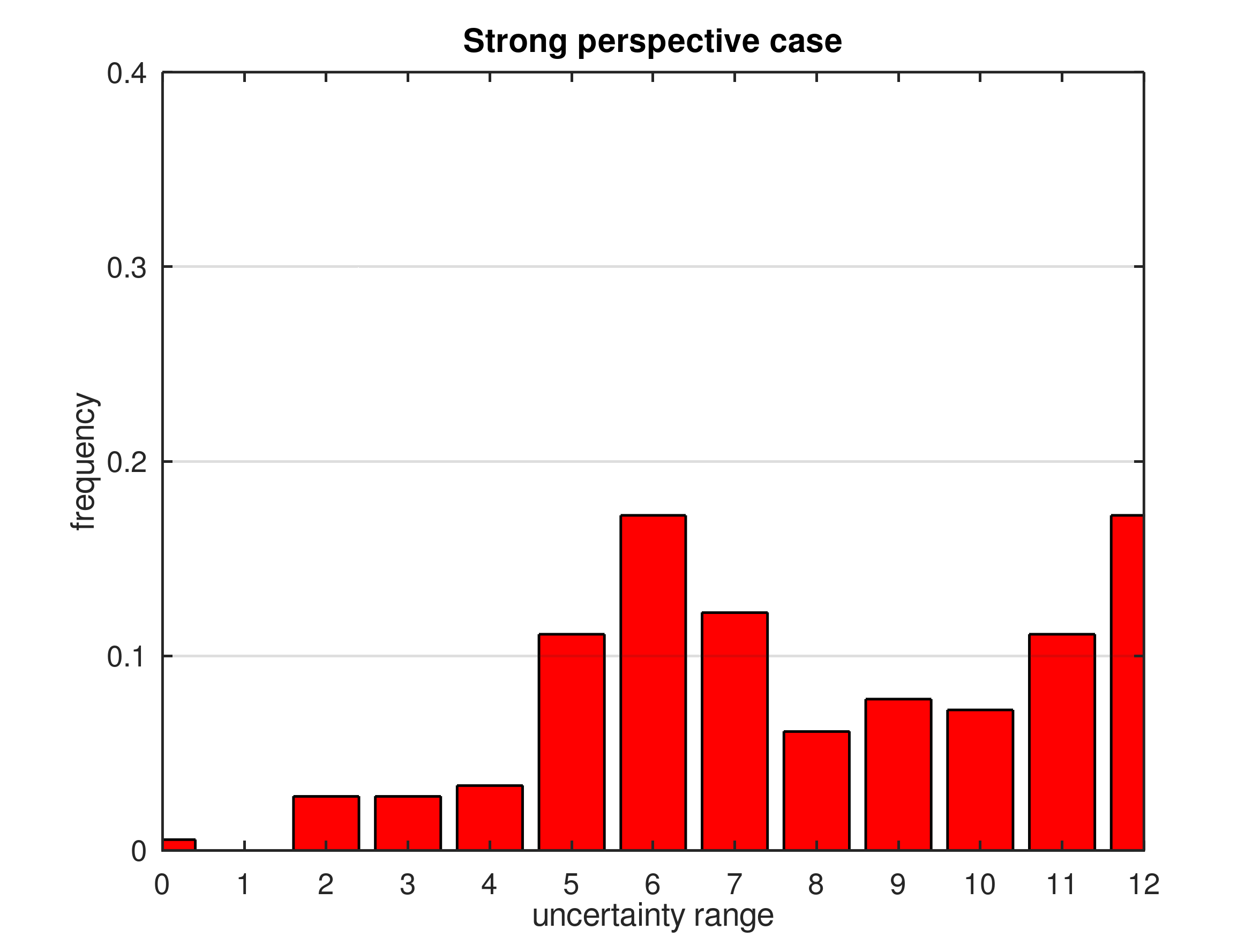}
  \caption{Strong perspective case}
  \label{fig:h2delta}
\end{subfigure}
\caption{Histograms of $\Delta v$ achieved in the speed estimates.}
\label{fig:hists_delta}
\end{figure}

As a second analysis, we examined the frequency with which the estimated speed either exactly matches the true speed or falls within small error margins ($\pm 1$, $\pm 2$ mph). These results are summarized in Table \ref{tab:result_perc}.
Notably, in $85\%$ of cases, the estimated speed deviates from the true value by less than $2$ mph, regardless of perspective.
\begin{table}[]
\begin{center}
\begin{tabular}{l|ccc}
Absolute error range  & exact speed & $\pm 1$ mph & $\pm 2$ mph \\
\hline
Low perspective case & 25\% (22)  & 61\% (54) & 84\% (75) \\
Strong perspective case & 25\% (45) & 71\% (127) & 85\% (153)
\end{tabular}
\caption{The columns report the percentage of speed estimates within a specific error range.
For instance, in the low perspective case, the absolute error is within $1$ mph the real speed in $54$ tests (representing $61\%$ of the cases)}
\label{tab:result_perc}
\end{center}
\end{table}
Additionally, we report the histogram of the uncertainty $\Delta v$ in the two scenarios.
In Figures \ref{fig:h1delta} and \ref{fig:h2delta}, the estimated values vary depending on the viewing angle.
Indeed, in most low-perspective cases, $\Delta v$ remains within a few mph.
In contrast, when the road appears at a steeper angle relative to the image plane, the estimated uncertainty range can become significantly larger, exceeding $12$ mph in some cases.
These results show the effectiveness of the proposed dataset in revealing the limitations of the analyzed speed estimation method. 
While the estimated speed generally aligns closely with the ground-truth values, despite the variety of camera settings and exporting settings, the method exhibits notable sensitivity to scene perspective.
In particular, under strong perspective distortion, the method tends to produce overly conservative uncertainty ranges that do not accurately reflect its actual performance.
In contrast, for scenes captured from low-perspective viewpoints, the method yields more reliable estimates with tighter and more representative confidence intervals.

\begin{figure}
\centering
\begin{subfigure}{.5\textwidth}
  \centering
\includegraphics[width=\linewidth]{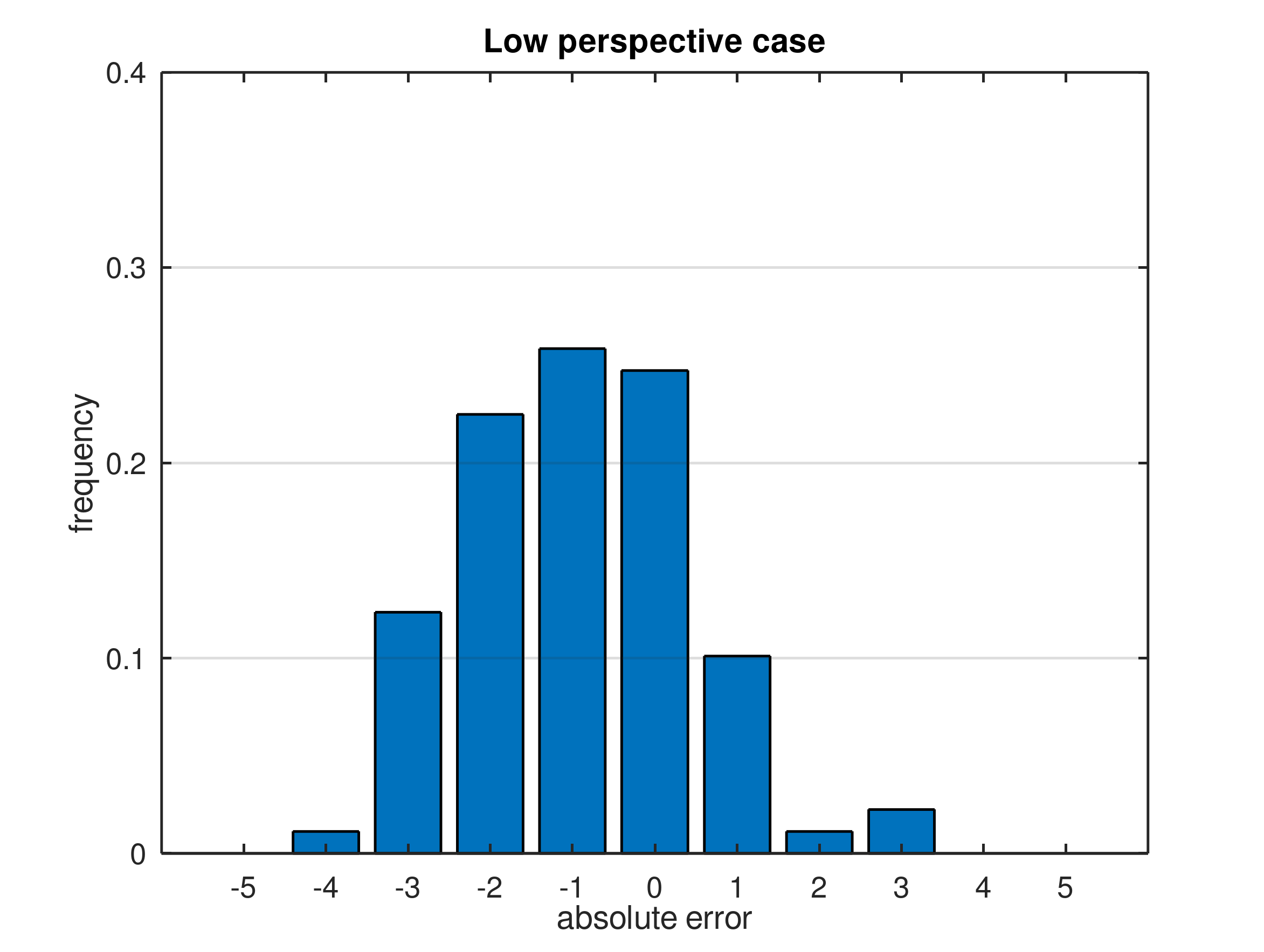}
  \caption{Low perspective case}
  \label{fig:h1}
\end{subfigure}%
\begin{subfigure}{.5\textwidth}
  \centering
\includegraphics[width=\linewidth]{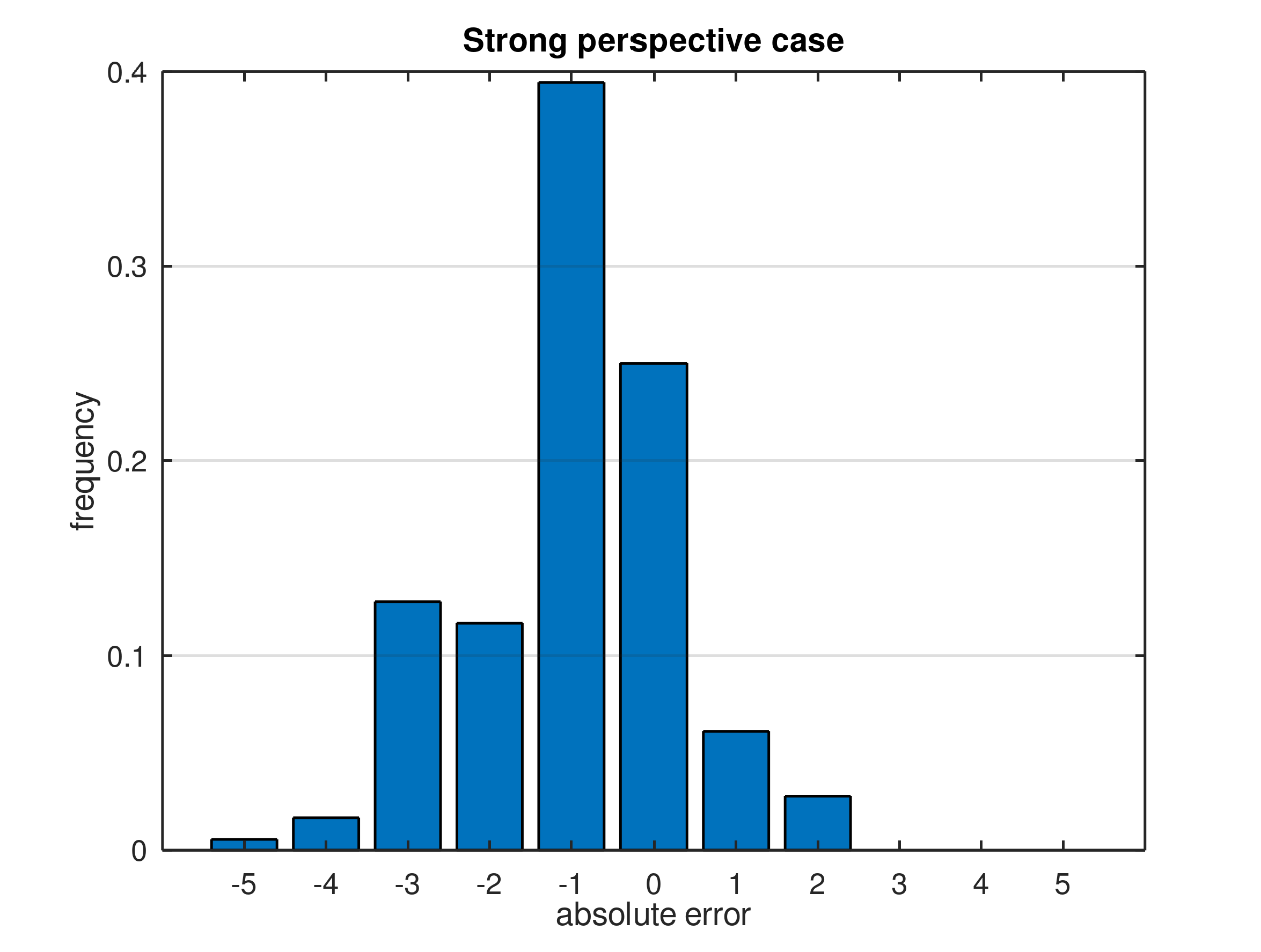}
  \caption{Strong perspective case}
  \label{fig:h2}
\end{subfigure}
\caption{Histograms of the absolute error between the estimated $v$ and the vehicle real speed.}
\label{fig:hists}
\end{figure}

\section{Conclusions}

We introduced ForeSpeed, a real-world dataset of $322$ CCTV camera videos.
The dataset is designed to evaluate the effectiveness of speed estimation methods over footage captured under varying acquisition conditions, including different scene perspectives and a range of spatial and temporal camera resolutions.
To resamble real scenarios, each video was extracted with various format and settings, based on the methods available on each camera.

As a case study, we employed ForeSpeed to assess the performance of a speed estimation method manually applied to the video sequences.
The results demonstrate that the method is generally capable of accurately estimating average speed, regardless of the scene perspective, camera settings, and exporting settings.
However, in scenarios involving strong perspective distortion, the method tends to produce overly broad uncertainty ranges.

ForeSpeed provides a valuable resource for exploring the limitations and specific characteristics of various speed estimation methods.
In the case of the method evaluated in this study, the dataset reveals a need for improvement in reducing output uncertainty under strong perspective distortion.

Moreover, ForeSpeed includes synchronized recordings of the same scene from multiple camera viewpoints, enabling future research on multi-view approaches to speed estimation.

\section*{Appendix - Dataset structure}
ForeSpeed comprises $269$ vehicle passes under CCTV cameras with various settings.
However, since the dataset is designed for forensic assessment, we take care of preserving the original video information that are relevant for the task, such as the presentation timestamps (PTS) of variable frame rate videos that are stored in the video container.
This led to some constraints in the file naming and splitting.
The dataset is structured as follow:
\begin{itemize}
    \item six folders, each containing the data acquired by a CCTV camera model (e.g., Eufy);
    \item each folder may contain multiple sub-folders if multiple cameras and storing formats are available (e.g.,  Lorex/cam1/main/avi). 
    These folders may include different CCTV cameras plugged into the same DVR, multiple quality strams that are available by the DVR, and multiple compression settings.    
    \item each sub-folder contains the $6$ and $8$ passes under the two considered perspectives. When possible, the video containing the pass is stored with the name $T<x>P<y>$ -$<$camera name$>$-$<$details$>$, where T1 and T2 stand for low and strong perspective respectively and P identify the pass (e.g., T1P5-Ring-6996443541347278144 stands for the pass $5$ with a low perspective in the camera Ring).
\end{itemize}
Since each camera uses a different acquisition protocol, some exceptions occurred:
\begin{itemize}
    \item Anran acquisition is split in four videos of $15$ minutes each;
    \item in some cameras the video acquisition is automatically split. When the split occurs during the vehicle pass, two videos of the same pass are included in the dataset (e.g., Lorex/cam1/main/avi/T1P4);
\end{itemize}
The number of the test and the pass is physically shown in the video before the pass. 
In Table \ref{tab:GT} we report the ground truth speed for each pass in each test.

\begin{table}[]
\caption{Ground truth speed in mph for each pass. T1 and T2 stands for low and strong perspectie respectively.}
\label{tab:GT}
\begin{center}
\begin{adjustbox}{width=0.8\textwidth}
\begin{tabular}{cccccc|cccccc|ccccc}
\multicolumn{5}{c}{ANRAN Cam 1}          &  & \multicolumn{5}{c}{ANRAN Cam 2}          &  & \multicolumn{5}{c}{EUFY}                 \\
Pass   & GT speed  &   & Pass & GT speed &  & Pass  & GT speed  &   & Pass  & GT speed &  & Pass  & GT speed  &   & Pass  & GT speed \\
\hline
T1P1   & 29        &   & T2P1 & 30       &  & T1P1  & 29        &   & T2P1  & 30       &  & T1P1  & 29        &   & T2P1  & 30       \\
T1P2   & 30        &   & T2P2 & 30       &  & T1P2  & 30        &   & T2P2  & 30       &  & T1P2  & 30        &   & T2P2  & 30       \\
T1P3   & 30        &   & T2P3 & 30       &  & T1P3  & 30        &   & T2P3  & 30       &  & T1P3  & 30        &   & T2P3  & 30       \\
T1P4   & 31        &   & T2P4 & 30       &  & T1P4  & 31        &   & T2P4  & 30       &  & T1P4  & 31        &   & T2P4  & 30       \\
T1P5   & 30        &   & T2P5 & 30       &  & T1P5  & 30        &   & T2P5  & 30       &  & T1P5  & 30        &   & T2P5  & 30       \\
T1P6   & 30        &   & T2P6 & 31       &  & T1P6  & 30        &   & T2P6  & 31       &  & T1P6  & 30        &   & T2P6  & 31       \\
       &           &   & T2P7 & 30       &  &       &           &   & T2P7  & 30       &  &       &           &   & T2P7  & 30       \\
       &           &   & T2P8 & 29       &  &       &           &   & T2P8  & 29       &  &       &           &   & T2P8  & 29       \\
       &           &   &      &          &  &       &           &   &       &          &  &       &           &   &       &          \\
\multicolumn{5}{c}{KASA}                 &  & \multicolumn{5}{c}{LOREX Cam 1 main avi} &  & \multicolumn{5}{c}{LOREX Cam 1 main mp4} \\
Pass   & GT speed  &   & Pass & GT speed &  & Pass  & GT speed  &   & Pass  & GT speed &  & Pass  & GT speed  &   & Pass  & GT speed \\
\hline
T1P1   & 29        &   & T2P1 & 30       &  & T1P1  & 29        &   & T2P1  & 30       &  & T1P1  & 29        &   & T2P1  & 30       \\
T1P2   & 30        &   & T2P2 & 30       &  & T1P2  & 30        &   & T2P2  & 30       &  & T1P2  & 30        &   & T2P2  & 30       \\
T1P3   & 30        &   & T2P3 & 30       &  & T1P3  & 30        &   & T2P3  & 30       &  & T1P3  & 30        &   & T2P3  & 30       \\
T1P4   & 31        &   & T2P4 & 30       &  & T1P4  & 31        &   & T2P4  & 30       &  & T1P4  & 31        &   & T2P4  & 30       \\
T1P5   & 30        &   & T2P5 & 30       &  & T1P5  & 30        &   & T2P5  & 30       &  & T1P5  & 30        &   & T2P5  & 30       \\
T1P6   & 30        &   & T2P6 & 31       &  & T1P6  & 30        &   & T2P6  & 31       &  & T1P6  & 30        &   & T2P6  & 31       \\
       &           &   & T2P7 & 30       &  &       &           &   & T2P7  & 30       &  &       &           &   & T2P7  & 30       \\
       &           &   & T2P8 & 29       &  &       &           &   & T2P8  & 29       &  &       &           &   & T2P8  & 29       \\
       &           &   &      &          &  &       &           &   &       &          &  &       &           &   &       &          \\
\multicolumn{5}{c}{LOREX Cam 1 sub mp4}  &  & \multicolumn{5}{c}{LOREX Cam 2 main avi} &  & \multicolumn{5}{c}{LOREX Cam 2 main mp4} \\
Pass   & GT speed  &   & Pass & GT speed &  & Pass  & GT speed  &   & Pass  & GT speed &  & Pass  & GT speed  &   & Pass  & GT speed \\
\hline
T1P1   & 29        &   & T2P1 & 30       &  & T1P1  & 29        &   & T2P1  & 30       &  & T1P1  & 29        &   & T2P1  & 30       \\
T1P2   & 30        &   & T2P2 & 30       &  & T1P2  & 30        &   & T2P2  & 30       &  & T1P2  & 30        &   & T2P2  & 30       \\
T1P3   & 30        &   & T2P3 & 30       &  & T1P3  & 30        &   & T2P3  & 30       &  & T1P3  & 30        &   & T2P3  & 30       \\
T1P4   & 31        &   & T2P4 & 30       &  & T1P4  & 31        &   & T2P4  & 30       &  & T1P4  & 31        &   & T2P4  & 30       \\
T1P5   & 30        &   & T2P5 & 30       &  & T1P5  & 30        &   & T2P5  & 30       &  & T1P5  & 30        &   & T2P5  & 30       \\
T1P6   & 30        &   & T2P6 & 31       &  & T1P6  & 30        &   & T2P6  & 31       &  & T1P6  & 30        &   & T2P6  & 31       \\
       &           &   & T2P7 & 30       &  &       &           &   & T2P7  & 30       &  &       &           &   & T2P7  & 30       \\
       &           &   & T2P8 & 29       &  &       &           &   & T2P8  & 29       &  &       &           &   & T2P8  & 29       \\
       &           &   &      &          &  &       &           &   &       &          &  &       &           &   &       &          \\
\multicolumn{5}{c}{Lorex Cam 2 sub mp4}  &  & \multicolumn{5}{c}{Lorex Cam 8 main avi} &  & \multicolumn{5}{c}{Lorex Cam 8 main mp4} \\
Pass   & GT speed  &   & Pass & GT speed &  & Pass  & GT speed  &   & Pass  & GT speed &  & Pass  & GT speed  &   & Pass  & GT speed \\
\hline
T1P1   & 29        &   & T2P1 & 30       &  & T1P1  & 29        &   & T2P1  & 30       &  & T1P1  & 29        &   & T2P1  & 30       \\
T1P2   & 30        &   & T2P2 & 30       &  & T1P2  & 30        &   & T2P2  & 30       &  & T1P2  & 30        &   & T2P2  & 30       \\
T1P3   & 30        &   & T2P3 & 30       &  & T1P3  & 30        &   & T2P3  & 30       &  & T1P3  & 30        &   & T2P3  & 30       \\
T1P4   & 31        &   & T2P4 & 30       &  & T1P4  & 31        &   & T2P4  & 30       &  & T1P4  & 31        &   & T2P4  & 30       \\
T1P5   & 30        &   & T2P5 & 30       &  & T1P5  & 30        &   & T2P5  & 30       &  & T1P5  & 30        &   & T2P5  & 30       \\
T1P6   & 30        &   & T2P6 & 31       &  & T1P6  & 30        &   & T2P6  & 31       &  & T1P6  & 30        &   & T2P6  & 31       \\
       &           &   & T2P7 & 30       &  &       &           &   & T2P7  & 30       &  &       &           &   & T2P7  & 30       \\
       &           &   & T2P8 & 29       &  &       &           &   & T2P8  & 29       &  &       &           &   & T2P8  & 29       \\
       &           &   &      &          &  &       &           &   &       &          &  &       &           &   &       &          \\
\multicolumn{5}{c}{Lorex Cam 8 sub mp4}  &  & \multicolumn{5}{c}{RING DevTools}        &  & \multicolumn{5}{c}{RING Download}        \\
Pass   & GT speed  &   & Pass & GT speed &  & Pass  & GT speed  &   & Pass  & GT speed &  & Pass  & GT speed  &   & Pass  & GT speed \\
\hline
T1P1   & 29        &   & T2P1 & 30       &  & T1P1  & 29        &   & T2P1  & 30       &  & T1P1  & 29        &   & T2P1  & 30       \\
T1P2   & 30        &   & T2P2 & 30       &  & T1P2  & 30        &   & T2P2  & 30       &  & T1P2  & 30        &   & T2P2  & 30       \\
T1P3   & 30        &   & T2P3 & 30       &  & T1P3  & 30        &   & T2P3  & 30       &  & T1P3  & 30        &   & T2P3  & 30       \\
T1P4   & 31        &   & T2P4 & 30       &  & T1P4  & 31        &   & T2P4  & 30       &  & T1P4  & 31        &   & T2P4  & 30       \\
T1P5   & 30        &   & T2P5 & 30       &  & T1P5  & 30        &   & T2P5  & 30       &  & T1P5  & 30        &   & T2P5  & 30       \\
T1P6   & 30        &   & T2P6 & 31       &  & T1P6  & 30        &   & T2P6  & 31       &  & T1P6  & 30        &   & T2P6  & 31       \\
       &           &   & T2P7 & 30       &  &       &           &   & T2P7  & 30       &  &       &           &   & T2P7  & 30       \\
       &           &   & T2P8 & 29       &  &       &           &   & T2P8  & 29       &  &       &           &   & T2P8  & 29       \\
       &           &   &      &          &  &       &           &   &       &          &  &       &           &   &       &          \\
\multicolumn{5}{c}{SWANN Cam 1 main avi} &  & \multicolumn{5}{c}{SWANN Cam 1 main mp4} &  & \multicolumn{5}{c}{SWANN Cam 1 sub avi}  \\
Pass   & GT speed  &   & Pass & GT speed &  & Pass  & GT speed  &   & Pass  & GT speed &  & Pass  & GT speed  &   & Pass  & GT speed \\
\hline
T1P1   & 29        &   & T2P1 & 30       &  & T1P1  & 29        &   & T2P1  & 30       &  & T1P1  & 29        &   & T2P1  & 30       \\
T1P2   & 30        &   & T2P2 & 30       &  & T1P2  & 30        &   & T2P2  & 30       &  & T1P2  & 30        &   & T2P2  & 30       \\
T1P3   & 30        &   & T2P3 & 30       &  & T1P3  & 30        &   & T2P3  & 30       &  & T1P3  & 30        &   & T2P3  & 30       \\
T1P4   & 31        &   & T2P4 & 30       &  & T1P4  & 31        &   & T2P4  & 30       &  & T1P4  & 31        &   & T2P4  & 30       \\
T1P5   & 30        &   & T2P5 & 30       &  & T1P5  & 30        &   & T2P5  & 30       &  & T1P5  & 30        &   & T2P5  & 30       \\
T1P6   & 30        &   & T2P6 & 31       &  & T1P6  & 30        &   & T2P6  & 31       &  & T1P6  & 30        &   & T2P6  & 31       \\
       &           &   & T2P7 & 30       &  &       &           &   & T2P7  & 30       &  &       &           &   & T2P7  & 30       \\
       &           &   & T2P8 & 29       &  &       &           &   & T2P8  & 29       &  &       &           &   & T2P8  & 29       \\
       &           &   &      &          &  &       &           &   &       &          &  &       &           &   &       &          \\
\multicolumn{5}{c}{SWANN Cam 1 sub mp4}  &  & \multicolumn{5}{c}{SWANN Cam 7 main avi} &  & \multicolumn{5}{c}{SWANN Cam 7 main mp4} \\
Pass   & GT speed  &   & Pass & GT speed &  & Pass  & GT speed  &   & Pass  & GT speed &  & Pass  & GT speed  &   & Pass  & GT speed \\
\hline
T1P1   & 29        &   & T2P1 & 30       &  & T1P1  & 29        &   & T2P1  & 30       &  & T1P1  & 29        &   & T2P1  & 30       \\
T1P2   & 30        &   & T2P2 & 30       &  & T1P2  & 30        &   & T2P2  & 30       &  & T1P2  & 30        &   & T2P2  & 30       \\
T1P3   & 30        &   & T2P3 & 30       &  & T1P3  & 30        &   & T2P3  & 30       &  & T1P3  & 30        &   & T2P3  & 30       \\
T1P4   & 31        &   & T2P4 & 30       &  & T1P4  & 31        &   & T2P4  & 30       &  & T1P4  & 31        &   & T2P4  & 30       \\
T1P5   & 30        &   & T2P5 & 30       &  & T1P5  & 30        &   & T2P5  & 30       &  & T1P5  & 30        &   & T2P5  & 30       \\
T1P6   & 30        &   & T2P6 & 31       &  & T1P6  & 30        &   & T2P6  & 31       &  & T1P6  & 30        &   & T2P6  & 31       \\
       &           &   & T2P7 & 30       &  &       &           &   & T2P7  & 30       &  &       &           &   & T2P7  & 30       \\
       &           &   & T2P8 & 29       &  &       &           &   & T2P8  & 29       &  &       &           &   & T2P8  & 29       \\
       &           &   &      &          &  &       &           &   &       &          &  &       &           &   &       &          \\
\multicolumn{5}{c}{SWANN Cam 7 sub avi}  &  & \multicolumn{5}{c}{SWANN Cam 7 sub mp4}  &  &       &           &   &       &          \\
Pass   & GT speed  &   & Pass & GT speed &  & Pass  & GT speed  &   & Pass  & GT speed &  &       &           &   &       &          \\
\hline
T1P1   & 29        &   & T2P1 & 30       &  & T1P1  & 29        &   & T2P1  & 30       &  &       &           &   &       &          \\
T1P2   & 30        &   & T2P2 & 30       &  & T1P2  & 30        &   & T2P2  & 30       &  &       &           &   &       &          \\
T1P3   & 30        &   & T2P3 & 30       &  & T1P3  & 30        &   & T2P3  & 30       &  &       &           &   &       &          \\
T1P4   & 31        &   & T2P4 & 30       &  & T1P4  & 31        &   & T2P4  & 30       &  &       &           &   &       &          \\
T1P5   & 30        &   & T2P5 & 30       &  & T1P5  & 30        &   & T2P5  & 30       &  &       &           &   &       &          \\
T1P6   & 30        &   & T2P6 & 31       &  & T1P6  & 30        &   & T2P6  & 31       &  &       &           &   &       &          \\
       &           &   & T2P7 & 30       &  &       &           &   & T2P7  & 30       &  &       &           &   &       &          \\
       &           &   & T2P8 & 29       &  &       &           &   & T2P8  & 29       &  &       &           &   &       &          \\
\end{tabular}
\end{adjustbox}
\end{center}
\end{table}

\bibliography{mybib}

\end{document}